\documentstyle[prd,aps,epsfig]{revtex}

\begin{document}
\topmargin -1.5cm \textheight 23cm
\newcommand{\beq}{\begin{equation}}
\newcommand{\eeq}{\end{equation}}
\newcommand{\beqn}{\begin{eqnarray}}
\newcommand{\eeqn}{\end{eqnarray}}
\newcommand{\slp}{\raise.15ex\hbox{$/$}\kern-.57em\hbox{$\partial
$}}
\newcommand{\slA}{\raise.15ex\hbox{$/$}\kern-.57em\hbox{$A$}}
\newcommand{\lnA}{\raise.15ex\hbox{$/$}\kern-.57em\hbox{$A$}}
\newcommand{\slB}{\raise.15ex\hbox{$/$}\kern-.57em\hbox{$B$}}
\newcommand{\bP}{\bar{\Psi}}
\newcommand{\bC}{\bar{\chi}}
\newcommand{\hs}{\hspace*{0.6cm}}
\title{Self-dual Ginzburg-Landau vortices in a disk }
 \author{
G.S.Lozano$^{a,b}$, M.V.Man\'{\i}as$^{c,b}$ and
E.F.Moreno$^{c,b}$}
\date{May 2000}
\address{\noindent $^a$ Depto. de F\'\i sica, FCEyN, Universidad de
Buenos Aires,  Pab I, Ciudad Univeristaria, Buenos Aires,
Argentina.\\ \noindent $^b$ Consejo Nacional de Investigaciones
Cient\'\i ficas y T\'ecnicas, Argentina. \\ \noindent $^c$  Depto.
de F\'\i sica. Universidad Nacional de La Plata.  CC 67, 1900 La
Plata, Argentina.}

\maketitle \thispagestyle{headings} \markright{\thepage}
\begin{abstract}
%\hs \vspace{4cm}

We study the properties of the Ginzburg-Laundau model in the
self-dual point for a two-dimensional finite system . By a
numerical calculation we analyze the solutions of the
Euler-Lagrange equations for a cylindrically symmetric ansatz. We
also study the self-dual equations for this case. We find that the
minimal energy configurations are not given by the Bogomol'nyi
equations but by solutions to the Euler Lagrange ones. With a
simple approximation scheme we reproduce the result of the
numerical calculation.
%\maketitle
\end{abstract}

\vspace{0.5cm}

Pacs:  11.27.+d,74,74.20.De,74.60.Ec

\maketitle

%\maketitle

%\newpage
%\pagenumbering{arabic} \normalsize \baselineskip 0.6cm

\vspace{1cm}

The study of vortex solutions in Ginzburg Landau (GL)
theories has been the subject of continuous interest in different
areas of condensed matter and high energy physics.

Static solutions in 2-dimensional infinite samples are
characterized  by the dimensionless GL parameter $\kappa$ which is
defined as the ratio of the magnetic penetration length $\lambda$
and the coherence length $\zeta$. It has been known for several
years that the GL model in the infinite plane possesses very
special properties at $\kappa^2=1/2$. For instance the second
order static equations of motion are equivalent to a set of much
simpler first order equations referred  as self-dual or
Bogomol'nyi equations (BE)\cite{B}.

Although the existence of self-dual equations was first pointed
out within  the study of superconductors by Harden and Arp
\cite{HA}  most of the research on the subject has been done in
connection with the role this type of equations play in High
Energy Physics. Indeed, self-dual equations were introduced in
this context independently by Bogomol'nyi and de Vega and
Schaposnik \cite{B} in their study of vortex solutions of the
relativistic version of the GL model (Abelian Higgs model).  Since
then, the properties of the solutions, the connection with
Supersymmetry, Topological Field Theories and Duality have been
established  not only for the Abelian Higgs model but also for
related theories in different number of space time dimensions and
non-Abelian gauge groups \cite{AG}.

Very recently, Akkermans and Mallick have addressed the study of
the GL model at the self-dual point for a 2-dimensional disk of
{\it finite} radius $R$ \cite{AM}. Their study is relevant for the
case of vortices in mesoscopic systems, where the size of the
sample is of the order of $\lambda$ and $\zeta$. As evidenced by
recent experiments \cite{Geim} , the superconducting behavior of
mesoscopic disks is radically different from that of the same
material in the macroscopic regime \cite{Pee}. An interesting
question that arises is then to determine in which way size
effects manifest at the self-dual point and reciprocally to
analyze if the special properties that the model show at the
self-dual point allow for a simpler interpretation of the
experimental results

In this work we shall re-analyze the properties of the GL model
for finite systems in the self-dual point. We will perform a
numerical study of the equation of motion and explore the role
played by the self-dual equations for this case. As a result of
our analysis it will be shown that some of the approximations made
in \cite{AM} are not correct. We shall in turn present a simple
approximation scheme which correctly reproduces the numerical
calculation.

The GL expression for the energy  of a 2-dimensional sample
$\Omega$ can be written as
\beq
 E = \int d^2x \left\{ \frac{1}{16 \pi} F_{ij}F_{ij} +\frac{1}{2}
|D_{i}\phi|^2 +V(|\phi|) \right\} \label{1} \eeq
where $F_{ij}=\partial_{i}A_{j}-\partial_{j}A_{i}$, and $
D_{i}\phi=\partial_{i}\phi -iq A_{i}\phi $ with $i=1,2$. Here
$A_i$ denotes the electromagnetic vector potential, $\phi$ is a
complex scalar field (order parameter) and $q$, the charge.
Writing the potential as
\beq V(|\phi|)= \beta/2 \left(|\phi|^2 - \eta^2\right)^2  \,\,
,\label{4} \eeq
the penetration length and the coherence length are
$\lambda^2=1/(4 \pi q^2 \eta^2)$ and $\zeta^2=1/(2 \eta^2 \beta)$
while the GL parameter is $\kappa^2=\beta/(2\pi q^2)$.

For arbitrary $\kappa$ the minimal energy configurations satisfy
the second order Euler Lagrange equations,
\beq D_{i}D_{i} \phi=-2\frac{\delta V}{\delta\phi^*} \label{5}
\eeq
\beq \frac{1}{4\pi}\partial_{i}F_{ij}=-j_{j}
=-q/2i(\phi^*D_{j}\phi-\phi D_{j}\phi^*) \label{6} \eeq
Using the identity
\beq \frac{1}{4}|D_{i}\phi \mp i \epsilon_{ij}D_j \phi|^2 =
\frac{1}{2} |D_{i}\phi|^2 \pm \frac{1}{2q} \epsilon_{ij}
\partial_i J_j
 \pm \frac{q}{2} B |\phi^2|
\eeq
the energy at the self-dual point can be re-written as
\beq {E} =\int d^2x \left( \frac{1}{16 \pi} (F_{ij}\pm q2\pi
\epsilon_{ij}(|\phi|^2-\eta^2) )^2
 +\frac{1}{4}
|D_{i}\phi \mp i \epsilon_{ij}D_j \phi|^2 \mp \frac{1}{2q}
\epsilon_{ij} \partial_i J_j \right) \mp \frac{q \eta^2}{2} \Phi
\label{r1}. \eeq
where $\Phi=\int_{\Omega} d^2x\; B = \oint_{\partial \Omega} {\vec
A}\cdot d{\vec x}$ is the total magnetic flux through the sample.
Assuming that the current is zero at the boundary, a lower bound
for the energy is obtained,
\beq E \geq |\frac{q \eta^2}{2} \Phi|.  \eeq
Energy configurations saturating the bound must  satisfy the
self-dual or Bogomol'nyi equations,
\beq F_{ij} \pm 2\pi q \epsilon_{ij}(|\phi|^2-\eta^2)=0 \; ,
\label{bogo1} \eeq
\beq D_{i}\phi \mp i \epsilon_{ij}D_j \phi=0. \label{bogo2} \eeq

In the plane these equations are totally equivalent to the Euler
Lagrange equations. To obtain finite energy conditions, one has to
demand
\beq \lim_{\rho\to\infty} D_i\Phi=0 \,\,\,\,\,\,
\lim_{\rho\to\infty} |\Phi|^2=\eta^2 \eeq
which in turns implies that
\beq
 \lim_{\rho\to\infty} J_i=0
 \label{ji};\eeq
On the other hand, writing $\phi=|\phi| e^{i\chi}$,  the current
takes the form:
\beq J_i=q|\phi|^2 (\partial_i \chi-q A_i). \eeq \label{ji2}
As $\phi$ is a single valued field, the phase $\chi(\rho,\theta)$
must satisfy
\beq \chi(\rho,2\pi)-\chi(\rho,0)=2\pi n, \label{chi} \eeq
with $n$ an integer. This condition together with (\ref{ji})
implies that the total flux has to be an integer multiple of the
quantum of the flux $\Phi_0=2\pi/q$
\beq \Phi=\int_{\partial \Omega} A_i
dx^i=\frac{1}{q}\int_{\partial \Omega}
\partial_i
 \chi dx^i= n\Phi_0 \;\; ,
\eeq
For each integer $n$, equations (\ref{bogo1}), (\ref{bogo2}) admit
a family of solutions depending on $2|n|$ parameters that can be
identified as the 2 dimensional coordinates of $|n|$ non
interacting vortices with flux quantum $\Phi_0$ (the upper or
lower sign in the equations has to be chosen according to the sign
of $n$)(\cite{jaf}).

Let us now concentrate to the case of finite geometries with a
boundary. When the current is zero at the boundary, the
Bogomol'nyi inequality still provides a bound for the energy and
the flux is quantized in units of $\Phi_0$. However, for a given
flux, there is a minimal area of the sample for which the
equations admit solutions. Indeed, integrating the first equations
we get the inequality:
\beq \Phi=\int_{\Omega}B=\int_{\Omega} 2\pi q (\eta^2-|\phi|^2)
\leq \int_{\Omega}2\pi \eta^2= \eta^2\frac{q}{2}\; area(\Omega).
\eeq

In the infinite plane the requirement of zero current at the
boundary is the natural boundary condition since it is the only
way of obtaining finite energy solutions. However there is no
compelling reason to do so for a finite region. The appropriate
boundary condition is \cite{SJ},
\beq D_{\bot} \phi=0\; . \label{bc} \eeq
Notice that this boundary condition only  implies vanishing of the
normal component of the current at the boundary. The tangential
component is left in principle undetermined. Nevertheless, it is
possible to show that for configurations satisfying the self-dual
equations, it holds the relation,
\beq J_i= \pm \frac{q}{2}\epsilon_{ij} \partial_j|\phi^2| \eeq
That is, for solutions of the BE equations both components of the
current must vanish at the boundary. As discussed above, this
implies that the total flux is quantized.

Suppose now that instead of imposing a boundary condition over
$J_{\|}$, we fix the total flux $\Phi$. It is clear, that if
$\Phi/\Phi_0$ is not an integer, a tangential component of the
current will be established at the boundary and minimal energy
configurations will not be given by solutions to the BE but by
solutions to the Euler Lagrange equations. More interestingly,
when $\Phi/\Phi_0=m$, with $m$ integer, although the BE do admit
solutions, they are not the minimal energy configuration. In fact,
it is energetically favorable to create $n \le m$ vortices and to
develop a tangential current at the boundary.

Let us illustrate this in a cylindrically symmetric ansatz:
\beqn \phi(x)& = & f(\rho)e^{in\theta}\nonumber\\ A_{\theta}(x)& =
& A(\rho) \nonumber \\ A_{\rho}(x)&=& 0\; . \label{8} \eeqn
Defining dimensionless variables  $x(r) = n-q A(r)$,
$z(r)=f(r)/\eta$, $r= (\rho/\lambda)$, the self-dual  equations
become
\beqn x'\pm \frac{r}{2}(z^2-1)&=&0 \label{16a}\\
z'\mp\frac{xz}{r} &=&0 \label{16b} \eeqn
and the boundary condition (\ref{bc}) translates into a Neumann
condition for the order parameter:
\beq z'(R^*)=0 \label{nbc}\eeq
where $R^*=R/\lambda$. Thus, we see that unless $x(R^*)=0$ (which
implies that $\Phi=\oint A=2\pi n/q$) the BE (\ref{16b}) cannot be
satisfied at $r=R^*$.

Minimal energy solutions are then obtained by solving the second
order Euler-Lagrange equations. In our ansatz they read,
\beq \frac{d^2x}{dr^2}-\frac{1}{r}\frac{dx}{dr}-xz^2=0 \label{12}
\eeq
\beq
\frac{d^2z}{dr^2}+\frac{1}{r}\frac{dz}{dr}-\frac{x^2z}{r^2}+\kappa^2
z(1-z ^2) =0 \label{13}, \eeq
while the energy can be expressed as,
\beq E=\pi \eta^2 \int_{0}^{R^*} r dr
[(\frac{x'}{r})^2+z'^2+\frac{x^2 z^2}{r^2}+\kappa^2 (z^2-1)^2]
\label{15} .\eeq
Regularity of cylindrical coordinates impose conditions at $r=0$,
and together with the Neumann condition (\ref{nbc}) and the
definition of $x(R^*)$ in terms of $n$ and $\Phi$ we have the
following set of boundary conditions:
\beqn x(0) & = & n \;\;\; x(R^*)=n-\frac{\Phi}{\Phi_0} \nonumber\\
z(0) & = & 0 \;\;\; z'(R^*)=0\; . \label{9} \eeqn

We have analyzed the existence of solutions of the system
(\ref{12},\ref{13},\ref{9}) by numerical integration. We employed
a relaxation method for boundary value problems \cite{NR}. In such
method, the differential equations are discretized in a convenient
mesh and converted to a set of coupled algebraic equations. The
system is then solved using Newton's iterative method, starting
from an initial guess and improving it iteratively. Given $\Phi$
and $R^*$ a solution for each integer $n$ is found corresponding
to a local minimum of the energy. Then, the $n$ giving the lowest
energy is selected.

In figure 1 we show the solutions corresponding to $\Phi=6 \Phi_0
,\;$  $R^*=10$. For these values, the minimal energy configuration
corresponds to $n=1$. The energy of this solution is $E=3.51 \pi
\eta^2$. As the external flux is an integer number of flux quanta,
the Bogomoln'nyi equations also have solutions for this case with
energy  $E=6 \pi \eta^2$. Thus, the system lowers its energy by
allowing a tangential component of the current $J_{||}(R^*) \neq
0$.

%%%%%%%%%%%%%%%%%%%%%%%%%%%%%%%%%%%%%%%%%
% Figure 1
\begin{figure}[ht]
\centerline{ \psfig{figure=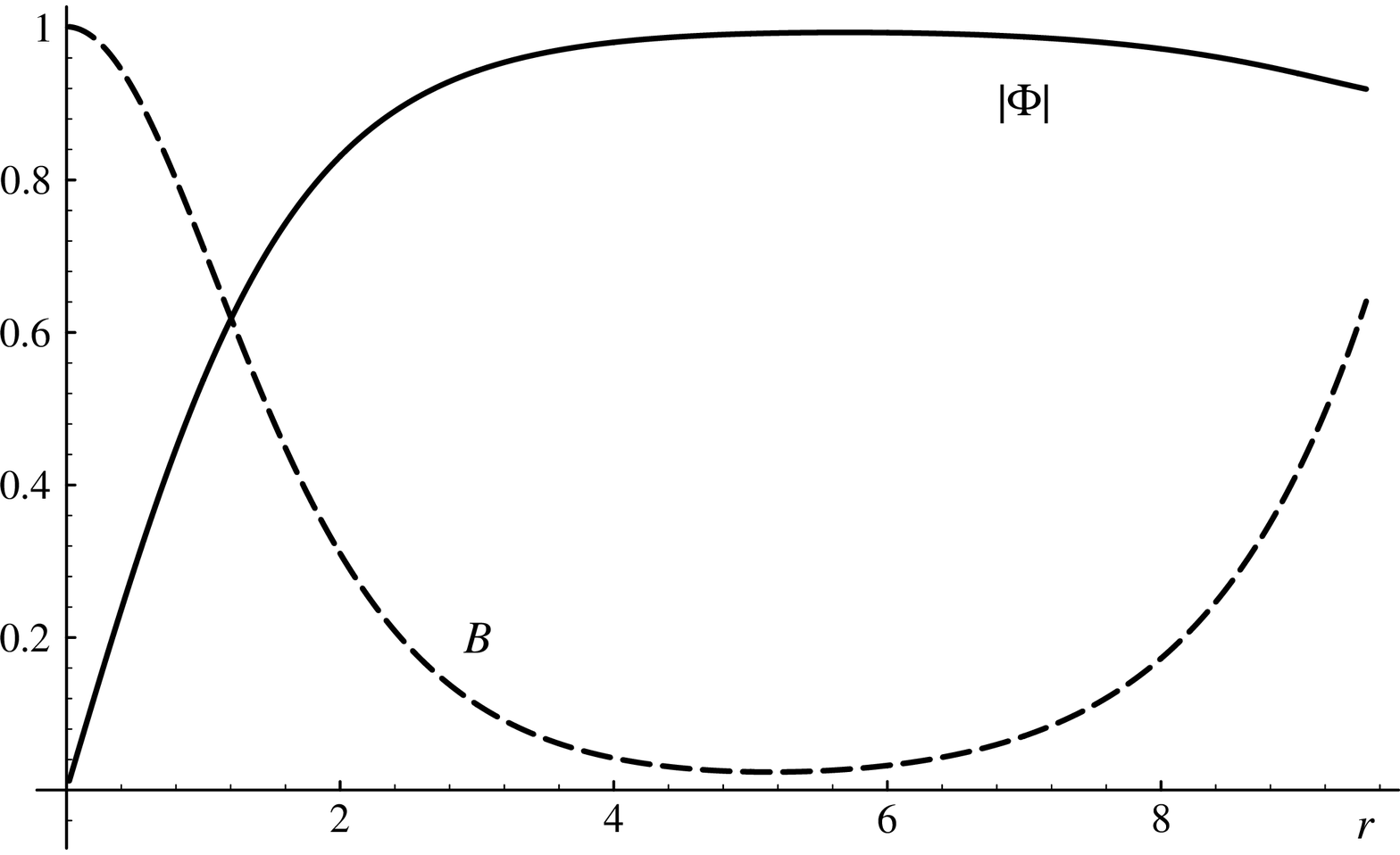,height=8cm,width=8cm,angle=0}
\psfig{figure=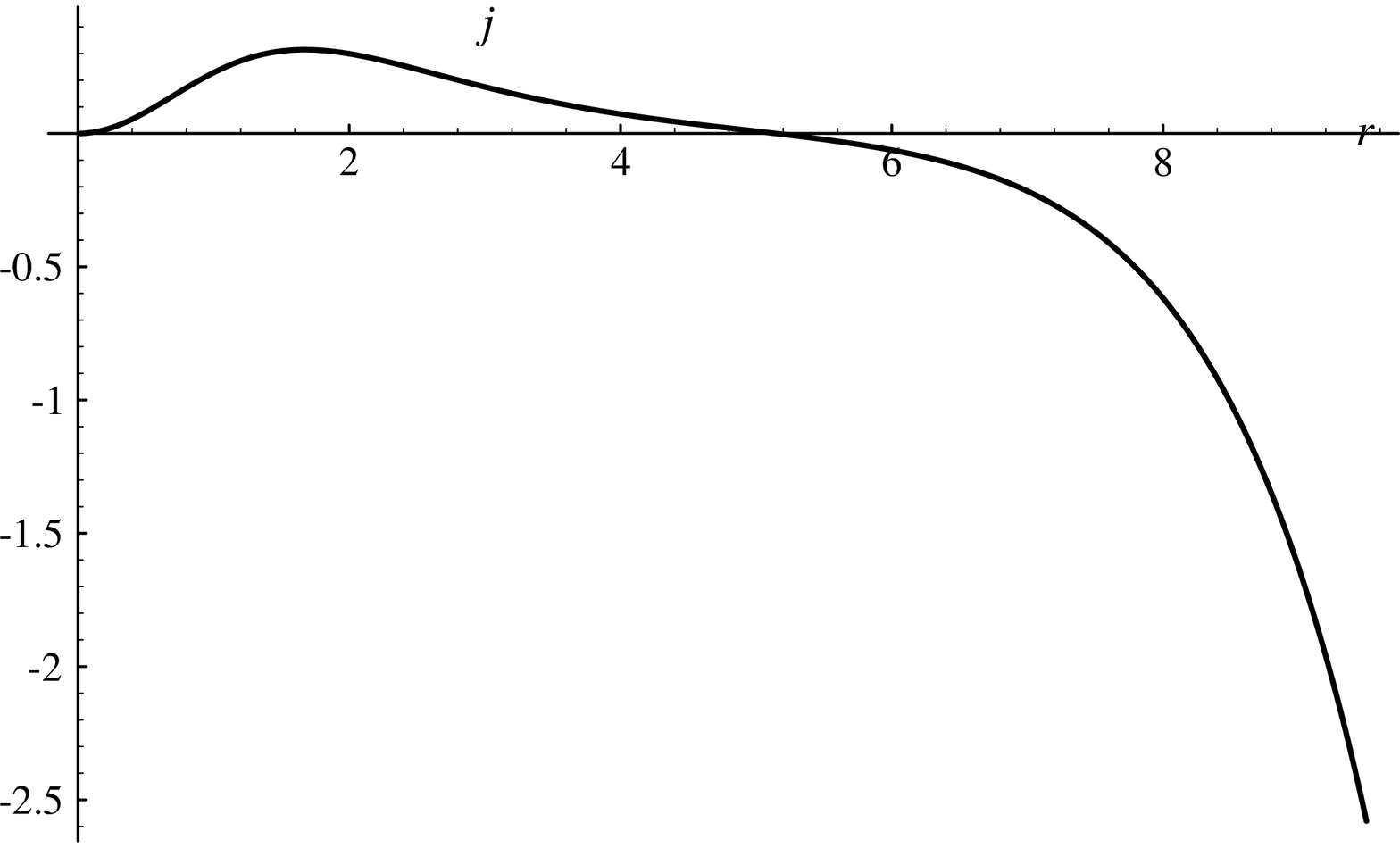,height=8cm,width=8cm,angle=0}}
\smallskip
\vspace{-2cm} \caption{ Vortex solution and angular component the
current $j(r)$ for
 $R=10$, $n=1$ and $\Phi=6 \Phi_0$.
 \label{plot1} }
\end{figure}
%%%%%%%%%%%%%%%%%%%%%%%%%%%%%%%%%%%%%%%%%

 Notice nevertheless that there is a point
$R_0^*$ such that $J(R_0^*)=0$. Following reference \cite{AM}, we
will consider the disk as $\Omega=\Omega_1 \cup \Omega_2$ where
$\Omega_1$, is the inner disk $ 0 \leq r\leq R_0^*$ and
$\Omega_2$, the outer ring $R_0^*\leq r \leq R^*$. We will express
the energy as:
\beq E(\Omega)=E(\Omega_1) + E(\Omega_2) \eeq

As the current vanishes at $R_0^*$, the authors in \cite{AM}
 assumed that  the minimal energy
configurations satisfy the Bogomol'nyi equations in $\Omega_1$;
however it  can be shown that this is not the case. Consider the
function $L[r]=z'(r)\mp x(r)\; z(r)/r$, evaluated with the
solutions of the Euler-Lagrange equations. If the Bogomol'nyi
equations are satisfied on $\Omega_1$, then $L[r]=0$ in this
region. But, as the current is not zero  on the external boundary
$r=R^*$, the Bogomol'nyi equations are not satisfied on
$\Omega_2$, and thus $L[r]\ne 0$ on $\Omega_2$. Clearly,
regularity of the solutions of ordinary differential equations
prevents the existence of a function that vanishes in the whole
region $0<r<R_0^*$ but is different from zero for $r>R_0^*$.
Although the solutions of the self-dual equation minimize the
energy on the internal region $\Omega_1$, any regular extension of
the solution to the whole disk will not minimize the total energy.

Having said this we should notice that even though the self-dual
solutions are not exact solutions in the inner disk $\Omega_1$,
they are in fact a very good approximation. A numerical analysis
of both solutions shows that for a wide range of parameters, they
differ only in around one part in a thousand and the same is true
for the energy. Then we will take,
\beq E(\Omega_1)\approx\pi \eta^2 |n| \eeq
Let us now analyze the contribution of the $\Omega_2$ region to
the energy.
\beq E(\Omega_2)=\pi \eta^2 \int_{R_0^*}^{R^*} r dr
[(\frac{x'}{r})^2+z'^2+\frac{x^2 z^2}{r^2}+\frac{1}{2} (z^2-1)^2]
\label{e2} \eeq
We first review the main steps in \cite{AM}. There it was assumed
that as the fields are concentrated in a region of width of order
one  from the border, this expression could be approximated as
\beq E(\Omega_2)\approx\pi \eta^2 \left. r
[(\frac{x'}{r})^2+z'^2+\frac{x^2 z^2}{r^2}+\frac{1}{2}
(z^2-1)^2]\right|_{r=R^*} . \label{e2.1}\eeq
Then, the condition $\frac{\delta E}{\delta z}=0$ would give
\beq \frac{x^2(R)}{R^2}=1-z^2 \label{con1} \eeq
As a next step Akkermans and Mallick neglected the magnetic energy
contribution (first term  in (\ref{e2.1})) arriving to the
following expression for the energy,
\beq E(\Omega_2)\approx\pi \eta^2 R^*
(\frac{x^2(R^*)}{R^{*2}}-\frac{1}{2} \frac{x^4(R^*)}{R^{*4}})
\label{e2.2}\eeq
Finally, the authors neglected, in the large $R$ limit, the
quartic term in (\ref{e2.2}) ending with the expression
\beq E(\Omega_2)\approx\pi \eta^2 \frac{x^2(R^*)}{R^{*}}
\label{e2.3} \eeq

The reasoning above suffers from two main drawbacks. Although the
field $z$ is practically constant, the field $x$ is not, making
the approximation of the integral not valid. In fact, our
numerical simulation shows that equation (\ref{con1}) is not
fulfilled. Second, as shown in fig 2 the magnetic energy is of the
same order than  the third term in (\ref{e2}).

%%%%%%%%%%%%%%%%%%%%%%%%%%%%%%%%%%%%%%%%%
% Figure 2
\begin{figure}[ht]
\centerline{ \psfig{figure=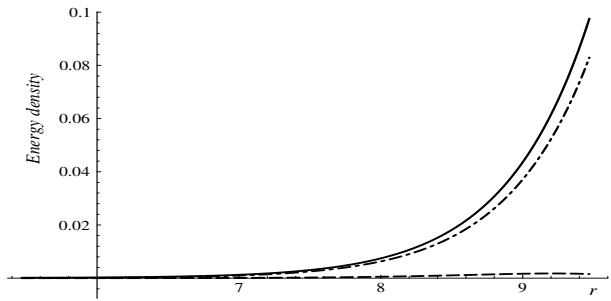,height=8cm,width=8cm,angle=0}}
\smallskip
\vspace{-2cm} \caption{ Energy density contributions near the edge
for various terms ($R^*=10, n=1$ and $\Phi=6 \Phi_0$). The filled
line corresponds to the term $B^2$, the dashed-dotted line
corresponds to the $|\phi|^2 |\nabla \chi - A|^2$ term , and the
dashed line to the term $|\nabla \Phi|^2$. Note that the $B^2$
term provides the largest contribution to the energy  and thus
cannot be neglected.}
\end{figure}
%%%%%%%%%%%%%%%%%%%%%%%%%%%%%%%%%%%%%%%%%
The correct field distribution in $\Omega_2$ can instead be
obtained by a simple approximation using
\beq \nu= \frac{(n-\frac{\Phi}{\Phi_0})}{R^*} \eeq
as an expansion parameter. Indeed, defining
\beq \tilde{x} = x /\nu \eeq
the equations of motion become
\beq \frac{d^2\tilde{x}}{dr^2}-\frac{1}{r}\frac{d\tilde{x}}{dr}-
z^2\tilde{x}=0 \label{21} \eeq
\beq
\frac{d^2z}{dr^2}+\frac{1}{r}\frac{dz}{dr}-\nu^2\frac{\tilde{x}^2
z}{r^2}+\frac{1}{2}z(1-z^2) =0 \label{22} \eeq
with the boundary conditions
\beq \tilde{x}(R_0^*)=0 \,\,\,\,\,\, \tilde{x}(R^*)=R^* \label{20}
\eeq
\beq z(R_0^*)=1 \,\,\,\,\,\, z'(R^*)=0 \eeq
In order to solve these equations, we make an expansion of the
form
\beqn \tilde{x}&=&\tilde{x}_0+\nu^2\tilde{x}_2+ O(\nu^{4})
\nonumber\\ z&=&1+\nu^2 z_2+ O(\nu^{4}) \label{23} \eeqn
The boundary conditions will be satisfied by the solution
$\tilde{x}_0$  while homogeneous conditions are valid for
$\tilde{x}_2$ and $z_2$. To lowest order we obtain
\beq
\frac{d^2\tilde{x}_0}{dr^2}-\frac{1}{r}\frac{d\tilde{x}_0}{dr}-
\tilde{x}_0=0 \label{24} \eeq
with solution
\beq \tilde{x}_0(r)=c_0
r\left[I_1(r)-\frac{I_1(R_0^*)}{K_1(R_0^*)}K_1(r)\right]
\label{31} \eeq
Here $c_0^{-1}=[I_1(R^*)-\frac{I_1(R_0^*)}{K_1(R_0^*)}K_1(R^*)]$
and $I_1(r)$, $K_1(r)$ are Bessel functions.

With these results we can solve the order $\nu^2$ equations:
\beq
\frac{d^2\tilde{x}_2}{dr^2}-\frac{1}{r}\frac{d\tilde{x}_2}{dr}-
\tilde{x}_2=2z_2\tilde{x}_0\; , \label{r32} \eeq
\beq \frac{d^2z_2}{dr^2}+\frac{1}{r}\frac{dz_2}{dr}-z_2
=\frac{\tilde{x}_0^2}{r^2} \label{r33} \eeq
with homogeneous boundary conditions for both functions. Using the
Green functions for each equation we obtain the solution
\beq
z_2(r)=y_2(r)\int_{R^*_0}^{r}\frac{y_1(r')f(r')}{W_{y_1,y_2}}dr'+
y_1(r)\int_{r}^{R^*}\frac{y_2(r')f(r')}{W_{y_1,y_2}}dr'\label{30}
\eeq
where
\beqn
y_1(r)&=&K_0(r)-\frac{K_0(R_0^*)}{I_0(R_0^*)}I_0(r)\nonumber\\
y_2(r)&=&K_0(r)-\frac{K'_0(R^*)}{I'_0(R^*)}I_0(r)\label{34}
\\ f(r)&=&\frac{\tilde{x}_0^2}{r^2} \nonumber \eeqn
and $W_{y_1,y_2}$ is the Wronskian.

The same steps are followed to calculate $\tilde{x}_2$ for which
we obtain
\beq
\tilde{x}_2(r)=Y_2(r)\int_{R_0^*}^{r}\frac{Y_1(r')g(r')}{W_{Y_1,Y_2}}dr'+
Y_1(r)\int_{r}^{R^*}\frac{Y_2(r')g(r')}{W_{Y_1,Y_2}}dr'\label{35}
\eeq
In this case
\beqn
Y_1(r)&=&r[K_1(r)-\frac{K_1(R_0^*)}{I_1(R_0^*)}I_1(r)]\nonumber\\
Y_2(r)&=&r[K_1(r)-\frac{K_1(R^*)}{I_1(R^*)}I_1(r)]\label{36} \\
g(r)&=&2z_2\tilde{x}_0 \eeqn
Having obtained the solutions
$\tilde{x}=\tilde{x}_0+\nu^2\tilde{x}_2$ and $z=1+\nu^2z_2$ we can
now consider the expression for the energy in this expansion:
\beqn E(\Omega_2)&=&\pi\eta^2\int_{R_0^*}^{R^*} dr \left\{
\frac{\nu^2}{r}[(\frac{d\tilde{x}_0}{dr})^2+\tilde{x}_0^2]+ \nu^4
[r(\frac{dz_2}{dr})^2+
\frac{2}{r}(\frac{d\tilde{x}_0}{dr}\frac{d\tilde{x}_2}{dr}+\tilde{x}_0^2z_2+
\tilde{x}_0\tilde{x}_2) \right. \nonumber  \\ &+& \left.
 r z_2^2]+ O(\nu^6) \right\} \label{37} \eeqn

Using the equations of motion, the energy takes the form

\beq \frac{E(\Omega)}{\pi\eta^2}= n +
\nu^2\frac{d\tilde{x}_0}{dr}(R^*)+ \nu^4\int_{R_0^*}^{R^*} dr
\frac{\tilde{x}_0^2 z_2}{r} \label{38} \eeq

The above  expression gives us the energy of the vortex
configuration up to the order $\nu^4$. However an obvious drawback
of this expression is the presence of the point $R_0^*$, which
should be located numerically, and then preventing any analytical
predictability power of the equation. Nevertheless we can convince
ourselves that the point $R_0^*$ can be shrunk to zero. The reason
is that at leading order we can approximate the whole solution as
a superposition of a Bogolmo'nyi vortex and a vortex concentrated
at the boundary. Because both kind of solutions are exponentially
small in complementary regions, any contribution from the
non-linearity of the equations is exponentially suppressed. In
fact, a simple plot of the solutions (\ref{31},\ref{30},\ref{35})
shows that the solution with $R_0^*=0$ only differs in about
$10^{-3}$ with the one with non-zero $R^*_0$. Therefore, we will
take $R^*_0=0$.

In this case the energy can be found by a simple numerical
integration. Notice that the integrals only depends on $R^*$ and
not on $\Phi$ or $n$. The result can be compared with the obtained
in reference \cite{AM}. In the following table we show a
comparison between the energy values obtained from our approximate
equation (\ref{38}), the ones obtained from the expression of
reference and \cite{AM} and the exact ones, corresponding to
vortex solutions with $R=10$ and $n=1$. We see that even for big
values of $\nu$, our approximate equation gives a result in
excellent agreement with the exact ones.

\vspace{.3 cm}

\centerline{
\begin{tabular}{|ccc|ccc|ccc|ccc|}
% after \\: \hline or \cline{col1-col2} \cline{col3-col4} ...
\hline
 ~ & ~ & ~ & ~ & ~ & ~ & ~ & ~ & ~ & ~ & ~ & ~\\
  ~ & $\Phi$ & ~ & ~ & {\rm Energy} & ~ & ~ & {\rm Energy}
 & ~ & ~ & {\rm Exact} &~\\
  ~ & ~ & ~ & ~ & {\rm from (\ref{38})} & ~ & ~ & {\rm of Ref.\cite{AM}}~
 & ~ & ~ & {\rm energy} &~\\
 ~ & ~ & ~ & ~ & ~ & ~ & ~ & ~ & ~ & ~ & ~ & ~\\\hline
 ~ & ~ & ~ & ~ & ~ & ~ & ~ & ~ & ~ & ~ & ~ & ~\\
 ~& 1.0 & ~ & ~ &  1.0 & ~ & ~ &
 1.0 & ~ & ~ &  1.0 &~\\ %\hline
  ~ & ~ & ~ & ~ & ~ & ~ & ~ & ~ & ~ & ~ & ~ & ~\\
 ~& 3.0 & ~ & ~ &  1.42 & ~ & ~ &
 1.4 & ~ & ~ &  1.42 &~\\ %\hline
 ~ & ~ & ~ & ~ & ~ & ~ & ~ & ~ & ~ & ~ & ~ & ~\\
 ~& 5.0 & ~ & ~ &  2.64 & ~ & ~ &
 2.6 & ~ & ~ &  2.64 &~\\ %\hline
 ~ & ~ & ~ & ~ & ~ & ~ & ~ & ~ & ~ & ~ & ~ & ~\\
 ~& 8.0 & ~ & ~ &  5.78 & ~ & ~ &
 5.9 & ~ & ~ &  5.79 &~\\ %\hline
  ~ & ~ & ~ & ~ & ~ & ~ & ~ & ~ & ~ & ~ & ~ & ~\\
 ~& 12.0 & ~ & ~ &  11.51 & ~ & ~ &
 13.1 & ~ & ~ &  11.27 &~\\ %\hline
 ~ & ~ & ~ & ~ & ~ & ~ & ~ & ~ & ~ & ~ & ~ & ~\\
 ~&18.0 & ~ & ~ &  18.81 & ~ & ~ &
 29.9 & ~ & ~ &  18.0 &~\\
 ~ & ~ & ~ & ~ & ~ & ~ & ~ & ~ & ~ & ~ & ~ & ~\\ \hline
\end{tabular}}

\vspace{.8 cm}

Furthermore we can find a large $R$ expansion of equation
(\ref{38}). Using the asymptotic expansion of the Bessel
functions, the first term in (\ref{38}) has the asymptotic form
$\nu^2 (R^* + 0.5 + O(1/R^*))$. The coefficient of  $\nu^4$ term
was found by  a numerical fit. The resulting approximate
expression for the energy is
\begin{equation}
\frac{E}{\pi\eta^2}= n+ \nu^2 R^* a(R^*) - \nu^4 R^* b(R^*) +
O(1/R^*, \nu^6)
\label{enn}
\end{equation}
where
\begin{equation}
a(R^*) = 1 + 1/(2 R^*) \; , \;\;\;\;\;\;\;\;\; b(R^*) = 0.139 +
0.111/R^*
\end{equation}

The most stable configuration corresponds to a vortex solution
with vortex number $n$ such that is an absolute minimum of the
energy. Equation (\ref{enn}) can be approximately minimized with
respect to $n$, with $b(R^*)/a(R^*)$ as the expansion parameter.
We find that the vortex number that minimizes the energy is given
by:
\begin{equation}
n =
%{\rm integer\;\;part}
\left[ \frac{\Phi}{\Phi_0}  - \frac{R^*}{2 a}
\left(\frac{a^3-b}{a^3 - 3/2 b}\right) + \frac{1}{2}\right]
\label{n}
\end{equation}
where $[x]$ means the {\it integer part of x}.

%%%%%%%%%%%%%%%%%%%%%%%%%%%%%%%%%%%%%%%%%
% Figure 3
\begin{figure}[ht]
\centerline{ \psfig{figure=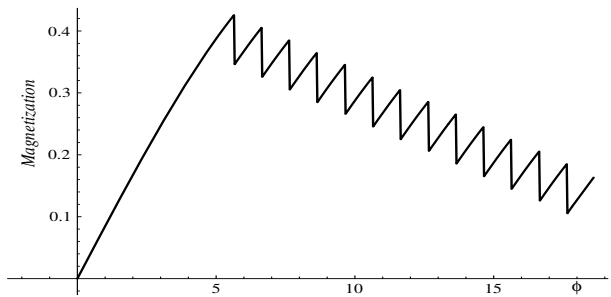,height=8cm,width=8cm,angle=0}}
\smallskip
\vspace{-2cm} \caption{ The magnetization $-M/(q \eta^2)$ as a
function of the external flux $\Phi/\Phi_0$ for $R^*=10$.}
\end{figure}
%%%%%%%%%%%%%%%%%%%%%%%%%%%%%%%%%%%%%%%%%

The magnetization of the system shown in fig.3 is obtained from
the Gibbs free energy, $G=E-2\pi \eta^2 (\Phi/\Phi_0)^2/R^{* 2}$,
as
\begin{equation}
M= -\frac{\partial G}{\partial \Phi} =- \frac{\pi \eta^2}{\Phi_0}
\left(\frac{2 a(R^*)}{R^*}\left( \frac{\Phi}{\Phi_0} - n\right) -
\frac{4 b(R^*)}{R^{* 3}}\left( \frac{\Phi}{\Phi_0} - n\right)^3 -
4 \frac{\Phi}{\Phi_0}\frac{1}{R^{* 2}} \right)
\end{equation}
where $n$ is given in equation (\ref{n}).

In this work we have analyzed the existence of vortex solutions of
the Ginzburg Landau theory at the self dual point $\kappa^2=1/2$
for a two dimensional disk of finite radius. Our original aim was
to pursue further the interesting proposal made by Akkermans and
Mallick {\cite{AM}} of exploiting the properties of the
Bogomol'nyi equations for the study of the vortices at the
self-dual point. Unfortunately our numerical study revealed that
some of the assumptions made there are not entirely correct.\\ We
have shown that the minimal energy configurations {\it do not}
satisfy the Bogomol'nyi equations in the inner disk. Nevertheless
they provide a very good approximation to the actual solutions.
Concerning the behaviour of the fields in the outer ring we have
provided a simple analytical approximation scheme which do conform
the numerical simulation and allows to obtain a selection rule for
the number of vortices as a function of the external flux.

%%%%%%%%%%%%%%%%%%%%%%%%%%%%%%%%%%%%%%%%%%%%%%%%%%%%%%%%%%%%%%%%%%%%%%%
%%%%%%%%%%%%%%%%%%%%%%%%%%%%%%%%%%%%%%%%%%%%%%%%%%%%%%%%%%%%%%%%%%%%%%%
%%%%%%%%%%%%%%%%%%%%%%%%%%%%%%%%%%%%%%%%%%%%%%%%%%%%%%%%%%%%%%%%%%%%%%%

%\begin{references}

%%%%%%%%%%%%%%%%%%%%%%%%%%%%%%%%%%%%%%%%%%%%%%%%%%%%%%%%%%%%%%%%%%%%%%%
%%%%%%%%%%%%%%%%%%%%%%%%%%%%  FIGURES   %%%%%%%%%%%%%%%%%%%%%%%%%%%%%%%
%%%%%%%%%%%%%%%%%%%%%%%%%%%%%%%%%%%%%%%%%%%%%%%%%%%%%%%%%%%%%%%%%%%%%%%

\end{document}